# Terahertz detection in a slit-grating-gate field-effect-transistor structure

D. M. Yermolayev,[1] K. M. Maremyanin,[2] D. V. Fateev,[3,4] S. V. Morozov,[2] N. A. Maleev,[4,5] V. E. Zemlyakov,[1] V. I. Gavrilenko,[2] S. Yu. Shapoval,[1] and V. V. Popov[3,4]

[1]*Institute of Microelectronic Technology and High-Purity Materials, Russian Academy of Sciences, Chernogolovka, Moscow Region 142432, Russia*

[2]*Institute for Physics of Microstructures, Russian Academy of Sciences, Nizhny Novgorod 603950, Russia*

[3]*Kotelnikov Institute of Radio Engineering and Electronics (Saratov Branch), Russian Academy of Sciences, Saratov 410019, Russia*

[4]*Saratov State University, Saratov 410012, Russia*

[5]*Ioffe Physical Technical Institute, Russian Academy of Sciences, St. Petersburg 194021, Russia*

**Abstract**

We have fabricated a grating-gate InGaAs/GaAs field-effect transistor structure with narrow slits between the grating gate fingers. The resonant photoconductive response of this structure has been measured in the sub-terahertz frequency range. The frequencies of the photoresponse peaks correspond to the excitation of the plasmon resonances in the structure channel. The obtained responsivity exceeds the responsivity reported previously for similar plasmonic terahertz detectors by two orders of magnitude due to enhanced coupling between incoming terahertz radiation and plasmon oscillations in the slit-grating-gate field-effect transistor structure.

Plasmonic response in the field-effect transistor with a two-dimensional (2D) electron channel can be used for detection of terahertz (THz) radiation [1]. Non-resonant [2-4] as well as the resonant [5-7] plasmonic detectors have been recently studied. The non-resonant plasmonic detectors are broadband THz detectors, which makes this type of detectors very promising for THz imaging [8-11]. The resonant (frequency selective) plasmonic detectors exhibit electrical tunability in a broad THz frequency band [7].

In a typical FET with 2D electron channel, the gate-to-channel separation is much smaller than the gate length $w$. In this case, the frequencies of the plasmon resonances excited in the FET channel under the gate with identical boundary conditions at its opposite ends are given by [1]

$$\omega = k_n \sqrt{\frac{eU_0}{m^*}}, \qquad (1)$$

where $U_0 = U_g - U_{th}$ is the difference between the gate voltage, $U_g$, and the channel depletion threshold voltage, $U_{th}$, and $e$ and $m^*$ are the electron charge and effective mass, respectively. The wavevectors of the gated plasmons, $k_n$, are quantized according the gate length as $k_n = (2n-1)\pi/w$ ($n = 1, 2, 3,…$).

One of the main problems to be solved with THz plasmonic devices is how to couple short-wavelength gated plasmon oscillations to relatively long-wavelength THz radiation (THz radiation wavelength typically exceeds the plasmon wavelength by two, or even three, orders of magnitude). The gated plasmons in a single-gate FET are weakly coupled to THz radiation [12] because: (i) the gated plasmons are strongly screened by the metal gate contact, (ii) they have a vanishingly small net lateral dipole moment due to their acoustic nature (in this mode, electrons oscillate out-of-phase in the gate contact and in the channel under the gate, which results in vanishingly small net lateral dipole moment), and (iii) gated plasmons strongly leak into ungated access regions of the channel. Therefore, special antenna elements are needed for coupling the plasmons in a 2D electron channel to THz radiation. A metal grating gate is an efficient coupler between plasmons in the FET channel and THz radiation [13]. The electromagnetic coupling between the plasmons and THz radiation considerably increases if the grating gate has narrow slits between its metal fingers. This is because strong near electric field excited at narrow slits strongly couples to plasmons in the grating-gate FET structure [14]. It is worth noting that, in a narrow-slit grating-gate FET structure, the selection rule for the excited plasmon modes is $k_n = 2\pi n/L$ ($n = 1, 2, 3,…$) [14, 15] rather than $k_n = (2n-1)\pi/w$ as for a single-gate FET structure [see Eq. (1)]. Well pronounced resonances at the fundamental plasmon frequency and its higher harmonics were observed in Ref. 15 in the narrow-slit grating-gate FET structure with strong electromagnetic coupling between plasmons and THz radiation. Resonant plasmonic THz detection was demonstrated recently in the grating-gate FET structures with the grating-gate duty cycle of 0.5 [16, 17]. The responsivity of these grating-gate FET plasmonic detectors still remains quite moderate (only few mV/W [17]). (Substantial enhancement in responsivity of the grating-gate FET detector can be achieved by placing it on a membrane substrate [18, 19] and by combining the grating-gate FET detector with on-chip micro-bolometer element involving a potential barrier under an isolated finger in the split-grating gate [18, 20].) For symmetry reasons, the grating-gate FET structure with a symmetric unit cell can exhibit only THz photoconductive plasmonic response which needs applying DC drain bias current in the structure channel. The THz photoconductive mechanisms in the grating-gate FET structures involve the plasmon-driven electron drag [21] and (in the channel with a spatially modulated electron density) the plasma electrostriction effect [22].

In this paper, we studied plasmonic THz photoconductive response of GaAs/InGaAs/AlGaAs grating-gate FET structure with a narrow-slit-grating gate. Heterostructure GaAs/InGaAs/AlGaAs was fabricated by the molecular-beam epitaxy (MBE) method. The 2D electron channel was formed in a 12-nm-thick undoped InGaAs layer with 40-nm-thick AlGaAs barrier layer which was $\delta$-doped with Si up to $5 \times 10^{12}$ cm$^{-2}$ and 400-nm-thick undoped GaAs buffer layer formed on the (100) surface of 450-m-thick semi-insulating (SI) GaAs substrate. The electron density in the channel was $3 \times 10^{12}$ cm$^{-2}$ with the electron effective mass $m^* = 0.061 m_0$, where $m_0$ is the free-electron mass, and room temperature mobility is 5900 cm$^2$/V·s. The grating-gate metallization was formed with 65-nm-thick Ti/Au/Ti by a standard lift-off process. The source and drain side metal contacts to the entire grating-gate FET structure (see the inset to Fig. 1) allowed for applying DC drain bias current in the structure channel and those contacts were also used for reading out the THz photoresponse signal. The grating-gate period was $L = 3$ $\mu$m with the width of each individual grating-gate finger $w = 2.7$ $\mu$m, i.e, the width of the slit between adjacent fingers of the grating gate was 0.3 $\mu$m (see the insets to Fig. 2). The gate-to-channel separation was 46 nm. All measurements were performed at liquid-helium temperature T = 4.2 K. The channel depletion threshold gate voltage obtained from the interpolation of the transfer characteristic of the grating-gate FET structure (Fig. 1) to the point of zero drain current, $I_d = 0$, was about $U_{th} = -1$ V. Figure 2 shows the I-V characteristic, which demonstrates the drain current saturation in the grating-gate FET structure channel for the source-to-drain voltages exceeding 0.8 V.

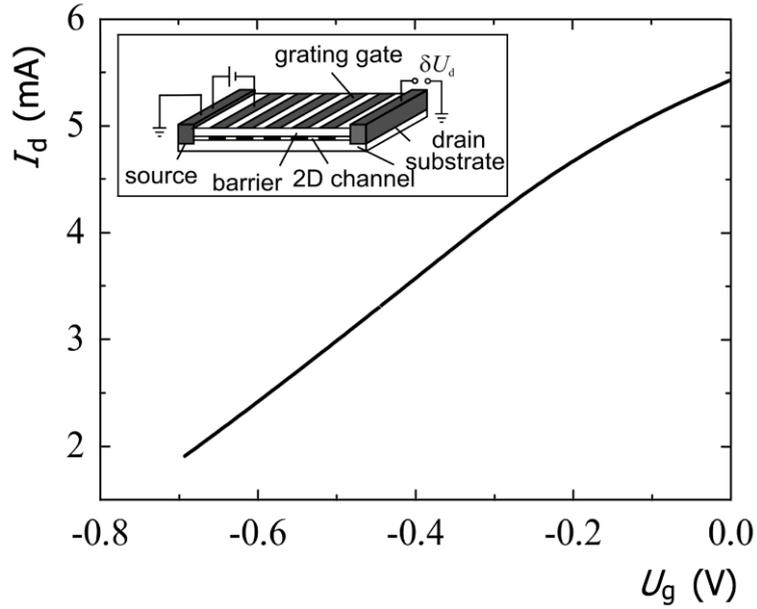

Fig. 1. Transfer characteristic of the GaAs/InGaAs/AlGaAs slit-grating-gate FET structure at temperature $T = 4.2$ K for the drain-to-source voltage $U_{ds} = 1$ V. Schematic view of the grating-gate-FET plasmonic THz detector is shown in the inset (THz radiation is incident from the top with the polarization of the electric field $E_0$ across the grating-gate fingers).

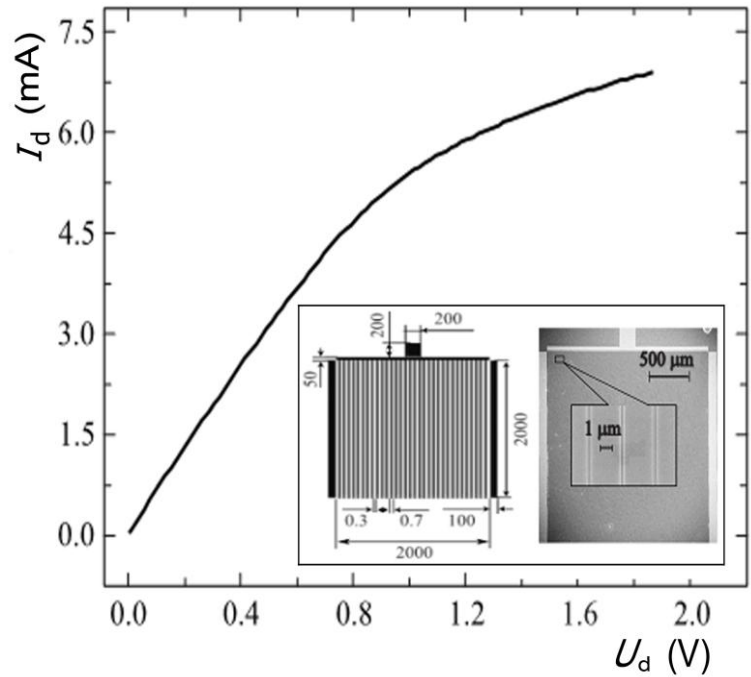

Fig. 2. I-V characteristic of the GaAs/InGaAs/AlGaAs slit-grating-gate FET structure at temperature $T = 4.2$ K for zero gate voltage $U_g = 0$. The inset shows the top metallization of grating-gate FET structure with all lateral dimensions designated (the left panel) and a scanning-electron-microscope image of the top view of the slit-grating gate FET (the right panel).

Large active area of the grating-gate FET structure ($2\times2$ mm$^2$) enabled efficient focusing of incident THz radiation upon the device area. Backward wave oscillator (BWO) was used as a monochromatic source of sub-THz radiation with the output power of about 1 mW in frequency range 0.415–0.72 THz. Terahertz radiation from the BWO was mechanically chopped at frequency 350 Hz and taken to the sample through an oversized circular metal waveguide with a tapered end which allowed focusing THz radiation onto the spot area of 6 mm in diameter covering the entire grating-gate FET structure. The electric field of the incident THz wave was polarized across the grating-gate fingers. Terahertz irradiation of the sample changes the channel resistivity $\rho_0$ by $\delta\rho$ which is the photoresistivity value. At fixed DC bias drain current $I_d = I_0$ (we chose $I_0 = 0.5$ mA which is much weaker than the saturation current), we measured the change of the voltage drop across the structure channel caused by THz irradiation of the sample (the photovoltage $\delta U_d$) as a function of the grating-gate voltage by a standard lock-in technique. In the fixed DC drain bias current regime, the photovoltage is proportional to the THz photoresistivity $\delta\rho$: $\delta U_d = I_0(\delta\rho)$.

Figure 3 (a) shows the channel photoresistivity as a function of the gate voltage for five different frequencies of the incoming THz radiation. For smaller frequencies, the peaks of the photoresistivity response appear at the gate voltages closer to the threshold voltage (which correspond to smaller $U_0$) in accordance with Eq. (1). Dividing the peak value of the photovoltage, $\delta U_{sd} \approx 4~\mu$V, by the THz power incident to the grating-gate FET area, we obtain the detection responsivity of 280 mV/W, which is two order of magnitude higher than was reported earlier for the grating-gate FET structure with the grating gate duty cycle of 0.5 [17]. Such strongly enhanced responsivity is ensured by strong coupling between incoming THz radiation and plasmons in the slit-grating-gate FET structure.

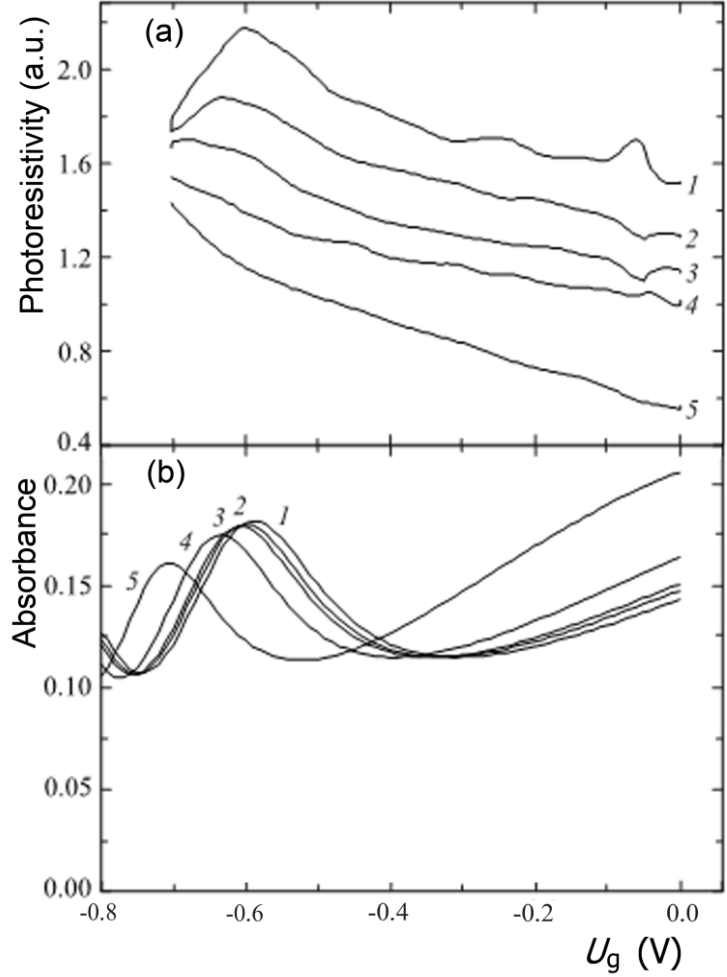

Fig. 3. (a) The THz photoresistivity response of the GaAs/InGaAs/AlGaAs slit-grating-gate FET structure at temperature $T = 4.2$ K as a function of the gate voltage for different frequencies: 697 GHz (curve 1), 688 GHz (curve 2), 682 GHz (curve 3), 659 GHz (curve 4), and 593 GHz (curve 5). Different curves corresponding to different frequencies are vertically shifted for clarity. (b) The calculated absorbance of the THz radiation in the GaAs/InGaAs/AlGaAs slit-grating-gate FET structure as a function of the voltage for the same THz frequencies as used in the THz photoconductive measurements.

Theoretical modeling of the interaction between the incident THz radiation and plasmons in the slit-grating-gate FET structure was performed in the self-consistent electromagnetic approach [23]. Profile of the equilibrium 2D electron density in the channel was calculated as a function of the gate voltages $U_g$ in the self-consistent electrostatic approach [24]. The only fitting parameter used in the modeling was the electron mobility in the FET structure channel, which was set at 30000 cm$^2$/V·s for the best matching the theoretical results with the experimental data. Figure 3 (b) shows the calculated absorbance of the THz radiation in the GaAs/InGaAs/AlGaAs slit-grating-gate FET structure as a function of the voltage for the five THz frequencies used in the THz photoconductive measurements. The absorption peaks correspond to the excitation of the second-order plasmon mode (with the plasmon wavevector $k_2 = 4\pi/L$) in the slit-grating-gate FET structure. The resonant absorbance reaches 0.2 which is close to the maximal theoretical value $A_{max} = 0.5\left(1 - \sqrt{R_0}\right)$, where $R_0$ is the reflectance of the substrate without the grating-gate FET structure [25]. This is the evidence of strong coupling between the incident THz radiation and the plasmon mode. Positions and lineshapes of the absorption resonances in Fig. 3 (b) match to those of the respective photoresistivity peaks in Fig. 2 (a). Therefore, we identify the observed THz photoresistivity peaks with the plasmonic response of the grating-gate FET channel. Resonant peaks of the photoresponse can hardly be seen in curved 4 and 5 in Fig. 3 (a) because they merge in a much stronger non-resonant photoresponse appeared for $U_g$ that are close to the channel depletion threshold [26].

In conclusion, we have demonstrated that the resonant plasmonic THz detection response can be considerably enhanced in the slit-grating-gate FET structure due to strong coupling between incident THz radiation and plasmon oscillations, which is ensured by narrow-slit grating gate. These results open up possibilities for improvement of the performance of the plasmonic THz detectors based on the grating-gate FET structures.

This work has been supported by the Russian Foundation for Basic Research (Grant Nos.10-02-93120 and 11-02-92101) and by the Russian Academy of Sciences Program Fundamentals of Nanotechnology and Nanomaterials". The work was performed under umbrella of the International Project "Semiconductor Sources and Detectors for Terahertz Frequencies."